\documentstyle[12pt]{article}

\begin{document}

\begin{center}
{\Large\bf
Glueball as a bound state in the self-dual homogeneous vacuum
gluon field}
\vspace*{.5cm}

{\bf Ja. V. Burdanov}, \\
Laboratory of Nuclear Problems,\\
Joint Institute for Nuclear Research, Dubna, Russia \\
{\bf G. V. Efimov}, \\
Bogoliubov Laboratory of Theoretical Physics, \\
Joint Institute for Nuclear Research, Dubna, Russia \\
\end{center}
\begin{abstract}
Using a simple relativistic QFT model of scalar fields
we demonstrate that the analytic confinement (propagator is an
entire function in the complex $p^2$-plane) and the weak
coupling constant lead to the Regge behaviour of the
two-particle bound states.

In QCD we assume that the gluon vacuum is realized by the self-dual
homogeneous classical field which is the solution of the Yang-Mills
equations. This assumption leads to analytical confinement
of quarks and gluons. We extract the colorless $0^{++}$ two-gluon
state from the QCD generating functional in the one-gluon exchange
approximation. The mass
of this bound state is defined by the Bethe-Salpeter equation.
The glueball mass varies in the region $1470{\it Mev}\leq
M_{0^{++}}\leq1600{\it Mev}$ for $0.2\leq\alpha_s\leq0.5$
if the gluon condensate is
${\alpha_s\over\pi}\langle G^a_{\mu\nu}G^a_{\mu\nu}\rangle=0.012
({\it Gev})^4$.
\end{abstract}

\newpage

\section{Introduction}

In this paper we shall calculate the mass of the
scalar glueball $O^{++}$. Before entering into a
discussion, formulation and solution of this
problem we would like to make some general remarks.

At the present time, QCD is considered to be an uniquely
true theory of strong interactions, but there are
different directions of its realization: lattice
calculations, QCD sum rules, Nambu-Yona-Lasinio model
plus or minus instantons, chiral perturbation theory,
calculation based on the Wilson loop,
bag and quasipotential models, models based on
the analytic confinement, and maybe something else.
It is useless to give references here because
the literature is so numerous that it is practicaly
impossible to list all the papers without missing
some of them.

In our opinion, theoretical reasons for this variety
consist in the following. From a general point of view
the QFT methods can be used if the propagators of
particles and vertices, which describe interactions,
are known and  coupling constants are {\it weak}.
Then perturbation calculations are valid and the
Bethe-Salpeter equation is an adequate relativistic
mathematical tool for computing bound states in QFT.
The quantum electrodynamics, where the propagators and
vertices are given and the coupling constant $\alpha={1\over137}$
is small, is the best example of this type of calculations.

However, in the case of QCD the situation is more difficult.
Although the QCD coupling constant is considered to be
relatively small $\alpha_s<1$, we can not be convinced
that propagators of quarks and gluons have the standard
Dirac and Klein-Gordon form for large distances, i.e.,
in the hadronization region. Unfortunately, a rigorous
generally accepted analytic solution of this problem
in QCD is still missing. Nevertheless, our hope is that a simple
formalism similar to QED should exist in QCD, but a background
form of these "basic bricks" is not yet found in QCD.

Another problem is that we do not have a self-consistent
conventional theory of bound states in QFT at all and
in QCD in particular, which can be applied in any cases
like the Schr\"{o}dinger equation is applied to any
nonrelativistic quantum phenomena. Therefore different approaches
with a doubtful basis are used to calculate bound states
in relativistic cases. One can see that in most cases
authors of different approaches strive to reduce the QFT bound state
problem to a nonrelativistic Schr\"{o}dinger equation with
an appropriate potential.

The generally accepted physical picture is that processes of
hadronization take place at large distances where the
confinement of quarks and gluons plays the main role. It should
lead to the following statements. First, the free Dirac and
Klein-Gordon equations for free particles are not adequate tools
to describe quarks and gluons in this region. Second,
currents, or vertices, which describe the connection of
quarks and gluons with hadrons, cannot be local.
In other words, only the knowledge of the vacuum QCD at
these distances should dictate what equations
have to describe quarks and gluons and nonlocal
vertices. In addition, the QCD effective coupling constant
$\alpha_s$ should be small enough in this hadronization region
if we want to use the Bethe-Salpeter type equations.

We think that a successful guess about the structure of QCD
vacuum at large distances can radically solve problems
of confinement and hadronization.

In this paper, we want to show, using a simple model of scalar
field, that in the frame work of QFT that analytic confinement of
constituent particles (propagators are entire
analytical functions in the momentum $p^2$-space) and
weak coupling constant (the Bethe-Salpeter equation
can be used) lead to the Regge spectrum of bound states.

Thus, if the QCD vacuum results in analytic confinement of
quarks and gluons and the QCD coupling constant $\alpha_s$
is small, hadron bound states are expected to have the
Regge type spectrum and we are able to calculate the masses
of these states using the Bethe-Salpeter equation.

In this paper, we calculate the mass of a glueball with quantum
numbers $0^{++}$. Glueballs are very interesting objects from
both theoretical and experimental points of view. The experimental
situation is not clear up to now (see, for example,
\cite{close,ellis}), and we shall deal with it.
However, the theoretical problem is quite intriguing: How
can massless gluons be joined together into massive objects
owing to the gluonic QCD self-interaction? It is evident
that the solution of the glueball problem is directly connected
with our understanding of nonperturbative Yang-Mills dynamics.
Theoretical interest in glueballs arised just from the
beginning of QCD and undoubtedly will continue until the
structure of QCD vacuum is understood.
All theoretical approaches (see, for example, the last papers
\cite{sim,forkel} where the previous publications are listed)
are based on the understanding that the structure of QCD vacuum
plays the decisive role in the solution of this problem, but all
authors avoid to make any direct guesses about its explicit
structure and take it into account in an indirect way only.

Our approach is based on the assumption that the QCD vacuum is
realized by the self-dual homogeneous vacuum gluon field which
is the classical solution of the Yang-Mills equations and
in the one-loop approximation the QCD vacuum energy has
minimum for nontrivial configurations of this field. This field
leads to the analytic confinement of quarks and gluons, so
that we could expect reasonable values for the glueball mass.
This vacuum gluon field has been applied to quarkonia in papers
\cite{mod1,mod2,mod3}. We shall use the functional integral
techniques to calculate mass spectrum of bound states (see
\cite{efim}). In this techniques the Bethe-Salpeter equation is
written in a symmetrized form permitting variational calculations.

\section{The toy model}

In this section, we consider a simple quantum field model with
confinement and demonstrate the features of possible bound states
which can be interpreted as standard physical particles.
We consider the scalar field $\varphi(x)$ with the Lagrangian in
the Euclidean metrics
\begin{eqnarray}
\label{Ltoy}
&&{\cal L}(x)=-{1\over2}\varphi(x)m^2e^{{\Box\over m^2}}
\varphi(x)-g\varphi^3(x).
\end{eqnarray}
The equation for the free field
$\phi(x)$ looks like
\begin{eqnarray}
\label{eqtoy}
&& m^2~e^{{\Box\over m^2}}\varphi(x)=0,~~~~~{\rm or}~~~~~
 m^2~e^{-{p^2\over m^2}}\tilde{\varphi}(p)=0
\end{eqnarray}
and the solution $\phi(x)\equiv0$ because the function
$e^{-{p^2\over m^2}}\neq0$, i.e., it has no zeroes for any
real or complex $p^2$. Exactly this property means analytic
confinement. The field $\phi(x)$ can exist in the virtual
state only, so that it can be called {\it virton field}
(see \cite{efiv}). In addition, one can say that the field
$\phi(x)$ describes constituent particles.

The propagator
\begin{eqnarray}
\label{prtoy}
&&\tilde D(p^2)={1\over m^2}~e^{-{p^2\over m^2}},~~~~~
D(x)=\int\!{dk\over(2\pi)^4}~\tilde D(p)~e^{ipx}=
{m^2\over(4\pi)^2}~e^{-{m^2\over4}x^2}.\nonumber\\
\end{eqnarray}
is an entire analytical function in the complex $p^2$-plane.
It guarantees the confinement of "particles" $\varphi(x)$
in each perturbation order over the dimensionless coupling
constant
\begin{eqnarray*}
&&\lambda=\left({3g\over 4\pi m}\right)^2\leq 1.
\end{eqnarray*}

The scale of the confinement region can be characterized by
the value
\begin{eqnarray}
\label{scale}
&& r_{cnf}^2={1\over\Lambda_{cnf}^2}={\int\!dx~x^2D(x)\over
\int\!dx~D(x)}={8\over m^2}.
\end{eqnarray}

The mechanism of arising of bound states can be described in
the following way. Let us consider the partition function
\begin{eqnarray}
&&Z=\int\!\delta\varphi~e^{{\cal L}[\varphi]}
=\int\!\delta\varphi~e^{-{1\over2}(\varphi D^{-1}\varphi)
+g\varphi^3}.
\nonumber
\end{eqnarray}
The effective Lagrangian with the one-particle exchange reads
\begin{eqnarray}
\label{Z}
&&{\cal L}_{eff}=-{1\over2}(\varphi D^{-1}\varphi)
+{(3g)^2\over2}(\varphi^2 D\varphi^2)+... .
\end{eqnarray}
The term $(\varphi^2 D\varphi^2)$ can be transformed as
\begin{eqnarray*}
L_2&=&{(3g)^2\over2}\int\!\!\!\int dx_1 dx_2~
\varphi^2(x_1) D(x_1-x_2)\varphi^2(x_2)\\
&=&{(3g)^2\over2}\int\!\!\!\int dx_1 dx_2~
\varphi(x_1)\varphi(x_2) D(x_1-x_2)\varphi(x_2)\varphi(x_1)\\
&=&{(3g)^2\over2}\int\!dx\int\!dy_1 \sqrt{D(y_1)}\int\!dy_2
\sqrt{D(y_2)}J(x,y_1)\delta(y_1-y_2)J^+(x,y_2),
\end{eqnarray*}
where $x_1=x+{1\over2}y,~~x_2=x-{1\over2}y$ and
\begin{eqnarray*}
&& J(x,y)=\varphi\left(x+{1\over2}y\right)\varphi
\left(x-{1\over2}y\right)=\varphi(x)~e^{{y\over2}\stackrel
{\leftrightarrow}{\partial}}\varphi(x),\\
&&~~~~~~~~~J^+(x,y)=J(x,-y).
\end{eqnarray*}

Let us introduce an orthonormal system $\bigl\{U_Q(y)\bigr\}$,
where $Q=(n,l,\{\mu\})$ are four-dimensional quantum numbers
so that
\begin{eqnarray}
\label{toyorto}
\int dy~U_Q(y)U_{Q'}(y)=\delta_{QQ'},~~~~{\rm and}
~~~~\sum_Q U_Q(y)U_Q(y')=\delta(y-y').
\end{eqnarray}
Then
\begin{eqnarray*}
&&L_2={(3g)^2\over2}\int\!dx~J_Q^+(x)\cdot J_Q(x),\\
&& J_Q(x)=\varphi(x)V_Q(\stackrel
{\leftrightarrow}{\partial})\varphi(x),~~~~~~
J_Q^+(x)=\varphi(x)V_Q(-\stackrel
{\leftrightarrow}{\partial})\varphi(x),\\
&&~~~~~~~~~~~~~V_Q(\stackrel{\leftrightarrow}{\partial})=
\int\!dy\sqrt{D(y)}U_Q(y)e^{{y\over2}\stackrel
{\leftrightarrow}{\partial}},
\end{eqnarray*}
and
\begin{eqnarray*}
&&e^{{(3g)^2\over2}(\varphi^2D\varphi^2)}=\exp\left\{
{(3g)^2\over2}\sum\limits_Q\int\!dx~J_Q^+(x)J_Q(x)\right\},\\
&&=\int d\sigma_{BB^+}~\exp\left\{{3g\over\sqrt{2}}\sum\limits_Q
\int\!dx~(B_Q^+(x)J_Q(x)+B_Q(x)J_Q^+(x))\right\}\\
&&=\int d\sigma_{BB^+}~\exp\left\{{3g\over\sqrt{2}}
\sum\limits_Q\int\!dx~\phi(x)[B_Q^+(x)V_Q(\stackrel
{\leftrightarrow}{\partial})+B_Q(x)V_Q(-\stackrel
{\leftrightarrow}{\partial})]\phi(x)\right\},
\end{eqnarray*}
where
$$ d\sigma_{BB^+}~=\prod_Q\delta B_Q~\delta B_Q^+~
e^{-\sum\limits_Q (B_Q^+B_Q)}.$$
We substitute this representation into the partition function
$Z$ and after integration over $\varphi$ we get
\begin{eqnarray}
\label{ZZ}
Z&=&\int\prod_Q\delta B_Q\delta B_Q^+~e^{-\sum\limits_Q(B_Q^+B_Q)
-{1\over2}{\rm Tr}\ln\left(1+3g\sqrt{2}
(B_Q^+V_Q+B_QV_Q^+)D+...\right)}\nonumber\\
&=&\int\prod_Q\delta B_Q\delta B_Q^+~
e^{-\sum\limits_{QQ'}\left(B_Q^+\left[\delta_{QQ\prime}-
\lambda\Pi_{QQ'}\right]B_Q\right)+O(B^3)}.
\end{eqnarray}
Here the polarization operator is
\begin{eqnarray*}
&&\lambda\Pi_{QQ'}(z)=(3g)^2{\rm Tr}~(V_Q^+DV_{Q'}D)\\
&&=\int\!\!\!\int dy_1dy_2~U_Q(y_1)\lambda\Pi(z;y_1,y_2)
U_{Q'}(y_2),\\
&& \lambda\Pi(z;y_1,y_2)=(3g)^2\sqrt{D(y_1)}
D\left(z+{y_1-y_2\over2}\right)D\left(z-{y_1-y_2\over2}
\right)\sqrt{D(y_2)}.
\end{eqnarray*}
where $z=x_1-x_2$.

The Fourier transform looks like
\begin{eqnarray*}
&&\lambda\tilde\Pi_{QQ'}(p)=\lambda\int\!dz~e^{ipz}\Pi_{QQ'}(z)
=\int\!\!\!\int\!dy_1dy_2~U_Q(y_1)S_p(y_1,y_2) U_{Q'}(y_2),
\end{eqnarray*}
whith
\begin{eqnarray}
\label{kernel}
&& S_p(y_1,y_2)=\lambda{m^4\over64\pi^2}e^{-{p^2\over2m^2}}
\cdot e^{-{m^2\over4}(y_1^2-y_1y_2+y_2^2)}.
\end{eqnarray}
The orthonormal system $\bigl\{U_Q(x)\bigr\}$ should diagonalize
the kernel $S_p(y_1,y_2)$
\begin{eqnarray*}
&&\lambda\tilde\Pi_{QQ'}(p)=\lambda_Q(p^2)\delta_{QQ'}.
\end{eqnarray*}
In other words, we have to find the spectrum and eigen functions
of the problem
\begin{eqnarray}
\label{BSeq}
&&S_p U_Q=\lambda_Q(p^2)U_Q~~~{\rm or}~~~
\int\! dy_2~S_p(y_1,y_2)U_Q(y_2)=\lambda_Q(p^2)U_Q(y_1).
\end{eqnarray}
It should be stressed that this equation is nothing else but the
Bethe-Salpeter equation in the one-boson exchange. In order to go to
the standard form of the Bethe-Salpeter equation, we have to
introduce new function $U_Q(y)=\sqrt{D(y)}\Psi_Q(y)$ and go
to the momentum representation (see, for example, \cite{efim}).

If we are able to do it, the partition function $Z$ in (\ref{ZZ})
looks like
\begin{eqnarray}
\label{ZB}
&&Z=\int\prod_Q\delta B_Q\delta B_Q^+~e^{-\sum\limits_Q\int\!dp
\left(\tilde{B}_Q^+(p)\left[1-\lambda_Q(p^2)\right]\tilde{B}_Q(p)
\right)+O(\tilde{B}^3)}
\end{eqnarray}
and the equation
\begin{eqnarray}
\label{Mass}
&&1-\lambda_Q(-M_Q^2)=0
\end{eqnarray}
defines the mass of the bound states with quantum numbers $Q$.
Then
\begin{eqnarray*}
&&1-\lambda_Q(p^2)=-\lambda^\prime_Q(-M_Q^2)(p^2+M_Q^2)+O((p^2
+M_Q^2)^2)
\end{eqnarray*}
and if
$$ -\lambda^\prime_Q(-M_Q^2)> 0 $$
we can interpret the field $B_Q$ as a physical particle with
quantum numbers $Q$ and mass $M_Q$.

The Bethe-Salpeter equation (\ref{BSeq}) can be solved
and the eigen functions look like
\begin{eqnarray*}
&& U_Q(y)=U_{nl\{\mu\}}(y)=C_{nl}T_{l\{\mu\}}(ay)L_n^{l+1}
(a^2y^2)e^{-{1\over2}a^2y^2},~~~~~a^2=m^2{\sqrt{3}\over2},
\end{eqnarray*}
Here, the angle polynomials $T_{l\{\mu\}}(y)=
T_{l\{\mu_1,...,\mu_l\}}(y)$ are normalized by
$$ \sum\limits_{\{\mu\}}T_{l\{\mu\}}(x)
T_{l\{\mu\}}(y)=C_{l}^1((xy))$$
where $C_{l}^1(t)$ are the Gegenbauer polynomials.
The eigenvalues are
\begin{eqnarray}
&& \lambda_{nl}(p^2)={\lambda~e^{-{p^2\over2m^2}}
\over(2+\sqrt{3})^{1+l+2n}}.
\end{eqnarray}
The equation (\ref{Mass}) gives the spectrum
\begin{eqnarray}
\label{sp-nl}
M^2_{nl}&=&2m^2\ln\left({\lambda_c\over\lambda}\right)+
(l+2n)\cdot2m^2\ln(2+\sqrt{3})
\end{eqnarray}
where $\lambda_c=(2+\sqrt{3})^2=13.928...$ .
One can see that this spectrum has the pure Regge behaviour.

Now we can represent
\begin{eqnarray*}
1-\lambda_Q(p^2)&=&1-{\lambda\over\lambda_c}\cdot
{e^{{M_Q^2\over2m^2}}\over(2+\sqrt{3})^{l+2n}}\cdot
e^{-{(p^2+M_Q^2)\over2m^2}}\\
&=&Z_Q(p^2+M_Q^2)+O((p^2+M_Q^2)^2)
\end{eqnarray*}
with
$$ Z_Q={\lambda\over\lambda_c}\cdot{e^{{M_Q^2\over2m^2}}
\over(2+\sqrt{3})^{l+2n}}\cdot{1\over2m^2}.$$
Thus, the kinetic term in $Z$ (\ref{ZB}) is
\begin{eqnarray*}
&&\left(\tilde{B}_Q^+(p)\left[1-\lambda_Q(p^2)\right]\tilde{B}_Q(p)
\right)\\
&&=Z_Q\left(\tilde{B}_Q^+(p)\left[(p^2+M_Q^2)+
O((p^2+M_Q^2)^2)\right]\tilde{B}_Q(p)\right).
\end{eqnarray*}
and after the renormalization
$$ \tilde{B}_Q(p)\to Z_Q^{-1/2}\tilde{B}_Q(p)$$
becomes of the standard form.

In particular, the lowest scalar bound state $B_0$ has the mass
\begin{eqnarray*}
&&M_{(00)}=m\sqrt{2\ln{\lambda_c\over\lambda}},
\end{eqnarray*}
For example, for $\lambda=.1$ we get
$$ M_{(00)}=3.142...\cdot m .$$

Finally, we conclude:
\begin{itemize}
\item Bound states exist for small coupling constants
$$\lambda < \lambda_c\approx 13 $$
and their masses grow when $\lambda\to0$ as
$$ M_Q\sim m\sqrt{2\ln{\lambda_c\over\lambda}}.$$

\item Analytic confinement leads to the pure Regge spectrum
for all bound states with quantum numbers $Q=(nl)$ and the
slope of the Regge trajectories is defined by the scale of the
confinement region $\Lambda_{cnf}^2$ and does not depend on the
coupling constant $\lambda$.

\item We can see that the size of the confinement region
$r_{cnf}$ (\ref{scale}) excels remarkably the Compton length
of all bound states
$$ l_Q={\hbar\over M_Qc}\ll r_{cnf}.$$
It means that the physical particles which are described by
the fields $B_Q(x)$ and all transformations with them take place
in the confinement region induced by the constituent particles
$\phi(x)$.

\item If $\lambda\leq .1$, i.e., we have the standard condition
for perturbation calculations, then
$$ M\ge m\sqrt{2\ln(10\lambda_c)}=3.142...\cdot m.$$
In other words, $M$ remarkably excels the "initial mass" $m$.
If $\lambda$ is small enough, the bound state can be quite large.

\end{itemize}

This scheme will be applied in main features to find the glueball
mass in QCD (see Table I).

\begin{tabular}{|c|c|}\hline
  & \\
  Toy model  & QCD \\
  & \\
\hline\hline
  & \\
 Gluon propagator is     &  QCD vacuum is realized \\
 assumed to be an entire & by self dual homogeneous\\
 analytic function     & gluon field $\breve{B}_\mu$\\
      $e^{-p^2}$         &  Gluon propagator is \\
                         &  an entire analytic function\\
 $\downarrow$   &   $\downarrow$    \\
 $L_I=g\phi^3(x) $         &  $ L_I=g{\rm Tr}(F_{\mu\nu}[A_\mu,A_\nu]) $  \\
 $\downarrow$   &   $\downarrow$    \\
 ${g^2\over2}\phi^2(x_1)D(x_1-x_2)\phi^2(x_2)$
       & ${g^2\over2}{\rm Tr}([A_\mu A_\nu]_{x_1}
       D_{\mu\nu,\alpha\beta}(x_1,x_2)[A_\alpha A_\beta]_{x_2})$\\
 $\downarrow$   &   $\downarrow$    \\
 ${g^2\over2}J(x_1,x_2)D(x_1-x_2)J(x_2,x_1),$
       & Firz transformation$\to$ scalar glueball \\
    & ${g^2\over2}{\cal J}(x_1,x_2){\cal W}(x_1-x_2){\cal J}(x_2,x_1)$\\
 $J(x,y)=\phi(x+y/2)\phi(x-y/2)$
       & ${\cal J}(x,y)=(A_\mu(x+y/2)e^{i\breve{b}(x,y)}A_\mu(x-y/2))$  \\
  $=\sum\limits_Q J_Q(x)U_Q(y)$
 $=\sum\limits_Q J_Q(x)U_Q(y) $ & \\
  $J_Q(x)=\phi(x)V_Q(\stackrel{\leftrightarrow}{\partial})\phi(x)$
  & ${\cal J}_Q(x)=(A_\mu(x)V_Q(\stackrel{\leftrightarrow}{\nabla})A_\mu(x))$  \\
$~~~~~~~\searrow$  & $\swarrow~~~~~~~$ \\
\multicolumn{2}{|c|}{}\\
 \multicolumn{2}{|c|}{$L_I={g^2\over2}\sum\limits_Q\int dx J_Q^2(x)$}\\
\multicolumn{2}{|c|}{$\downarrow$}\\
\multicolumn{2}{|c|}{$e^{{g^2\over2}(J_Q^2)}=\int DB_Q~e^{-(B_Q^2)+g(B_QJ_Q)}$ and
 Integration over $\phi(x)$ and $A_\mu(x)$}\\
\multicolumn{2}{|c|}{$\downarrow$}\\
\multicolumn{2}{|c|}{$\int DB_Q~ e^{-(B_Q^2)-{\rm Tr}\ln[1+g(B_QV_Q)D]}\to
 \int DB_Q ~e^{-(B_Q[1-g^2{\rm Tr}(V_QDV_QD)]B_Q)+O(B_Q^3)}$}\\
\multicolumn{2}{|c|}{$\downarrow$}\\
\multicolumn{2}{|c|}{Bethe-Salpeter equation:}\\
\multicolumn{2}{|c|}{$g^2{\rm Tr}(V_QDV_{Q'}D)=(U_Q\Pi U_{Q'})=\lambda_Q(p^2)\delta_{QQ'}$}\\
\multicolumn{2}{|c|}{$\downarrow$}\\
\multicolumn{2}{|c|}{Mass spectrum: $1=\lambda_Q(-M_Q^2)$ }\\
\multicolumn{2}{|c|}{}\\
\hline
\end{tabular}

\section{QCD Lagrangian in the self-dual gluon field}

Now let us go to QCD. The action has the form
\begin{eqnarray}
\label{QCD}
&&S=-\frac{1}{2}\int {\rm d}^4x~{\rm Tr}~\hat G_{\mu \nu}
\hat G_{\mu \nu},\\
&&\hat G_{\mu \nu}=\partial_\mu\hat A_\nu-\partial_\nu\hat A_\mu
-ig[\hat A_\mu,\hat A_\nu],~~~~~\hat A_\mu=t^a A_\mu^a.\nonumber
\end{eqnarray}
We suppose (see \cite{mod1,mod2,mod3}) that the QCD vacuum is
realized by a homogeneous (anti-)self-dual gluon field which has
the form
\begin{eqnarray}
\label{B}
&& \hat{B}_\mu(x)=\hat{B}_{\mu\nu}x_\nu=
\hat{n}Bb_\mu(x)=\hat{n}Bb_{\mu\nu}x_\nu~,\\
&&\hat n=t^an^a,~~~~~\hat{n}^+=\hat{n},~~~~~(nn)=n^an^a=1,
\nonumber\\
&& b_{\mu\rho}b_{\rho\nu}=-\delta_{\mu\nu},~~~~~~~
\tilde{b}_{\mu\nu}=\epsilon_{\mu\nu\alpha\beta}b_{\alpha\beta}
=\pm b_{\mu\nu}.\nonumber
\end{eqnarray}
After substitution
\begin{eqnarray*}
&&~~~~~~~~~~~~\hat A_\mu(x)\to \hat A_\mu(x)+\hat B_\mu(x)~,
\end{eqnarray*}
one has
\begin{eqnarray*}
\hat G_{\mu \nu}&=&\hat{\cal F}_{\mu\nu}
-ig[\hat A_\mu,\hat A_\nu]-2\hat B_{\mu\nu}.
\end{eqnarray*}
Here
\begin{eqnarray*}
&&\hat{\cal F}_{\mu\nu}=\breve\nabla_\mu A_\nu
-\breve\nabla_\nu A_\mu,
\end{eqnarray*}
where we have used the notation
\begin{eqnarray*}
&& \breve{\nabla}_\mu^{ab}=\breve{\nabla}_\mu^{ab}(x)=
\delta^{ab}{\partial\over\partial
x_\mu}-i\breve{\Omega}^{ab}b_\mu(x),~~~~~
\breve\nabla_\mu={\partial\over\partial x_\mu}
-i\breve{\Omega}b_{\mu\nu}x_\nu,\\
&&~~~~~~[\breve{\nabla}_\mu,\breve{\nabla}_\nu]=2i\breve{\Omega}
b_{\mu\nu},~~~~~~\breve{\Omega}=gB\breve{n},~~~~~~
\breve{\Omega}^{ab}=gB\breve{n}^{ab},
\end{eqnarray*}
$$\breve n^{ab}=-if^{abc}n^c,~~~~~~
\breve{n}^+=\breve{n},~~~~~~\breve{n}^\top=-\breve{n}.$$

The QCD action (\ref{QCD}) in the background field $B$ looks like
\begin{eqnarray*}
S&=&-\frac{1}{2}\int {\rm d}^4x~{\rm Tr}~
\left\{\hat{\cal F}_{\mu\nu}\hat{\cal F}_{\mu\nu}+
4ig\hat B_{\mu\nu}[\hat A_\mu,\hat A_\nu]\right.\\
&-&\left.2ig\hat{\cal F}_{\mu\nu}[\hat A_\mu,\hat A_\nu]
+4\hat B_{\mu\nu}\hat B_{\mu\nu}-
g^2[\hat A_\mu,\hat A_\nu]^2\right\}.
\end{eqnarray*}
In order to fix a gauge, we should add to the action
the gauge-fixed term
\begin{eqnarray}
S_{gf}=\int {\rm d}^4x~{\rm Tr}~
(A_\mu\breve\nabla_\mu)(\breve\nabla_\nu A_\nu)=
-\frac{1}{2}\int {\rm d}^4x~(\breve\nabla_\mu^{ab} A_\mu^b)^2.
\nonumber
\end{eqnarray}
After some transformations the QCD action is
\begin{eqnarray*}
S+S_{gf}&=&\int {\rm d}^4x~{\rm Tr}~\left\{(A_\mu\breve
\nabla_\nu)(\breve\nabla_\nu A_\mu)+8ig\hat A_\mu
\hat B_{\mu\nu}\hat A_\nu\right\}\\
&+&\int {\rm d}^4x~{\rm Tr}~\left\{ig\hat{\cal F}_{\mu\nu}
[\hat A_\mu,\hat A_\nu]+\frac{g^2}{2}[\hat A_\mu,
\hat A_\nu]^2-2\hat B_{\mu\nu}\hat B_{\mu\nu}\right\},
\end{eqnarray*}

The part of action quadratic over the quantum fields $A_\mu^a$
can be reduced to the form
\begin{eqnarray}
&&\int {\rm d}^4x~{\rm Tr}~\left[(A_\mu\breve\nabla_\nu)
(\breve\nabla_\nu A_\mu)+8ig\hat A_\mu\hat B_{\mu\nu}\hat
A_\nu\right]=-\frac{1}{2}\int {\rm d}^4x~
A_\mu^a(G^{-1})^{ab}_{\mu\nu}A_\nu^b,\nonumber
\end{eqnarray}
where
\begin{eqnarray}
\label{Grgl}
&&(G^{-1})^{ab}_{\mu\nu}=\left[-\breve{\nabla}^2_\rho
\delta_{\mu\nu}-4i\breve{\Omega}b_{\mu\nu}\right]^{ab}.
\end{eqnarray}

Finally, the QCD action reads
\begin{eqnarray}
\label{QCDaction}
S&=&S_0+S_I,\\
S_0&=&-{1\over2}\int dx~A_\mu^a(G^{-1})_{\mu\nu}^{ab}
A_\nu^b=-{1\over2}~(A_\mu G^{-1}_{\mu\nu}A_\nu),\nonumber\\
S_I&=&\int {\rm d}^4x~{\rm Tr}~\left\{ig
\hat{\cal F}_{\mu\nu}[\hat A_\mu,\hat A_\nu]+
\frac{g^2}{2}[\hat A_\mu,\hat A_\nu]^2\right\}.\nonumber
\end{eqnarray}

\section{The Gluon Green function}

Let us find the gluon Green function corresponding to
the operator (\ref{Grgl}). The $8\times8$ matrix $\breve{\Omega}$
for any vector $n^a$ has at least two zeroth eigenvalues.
The choice of a vector $n^a$ requires special consideration
and we plan to do it in further publications. In this paper,
we choose the simplest case $n^8=1,~~~n^a=0$ for $a=1,...,7$
which guarantees the quark confinement (see \cite{mod1,mod2,mod3}).
In this case, it is convenient to introduce the notion
$$ \breve{e}={\rm diag}(1,1,1,0,0,0,0,1),~~~~~~~
\breve{h}={\rm diag}(0,0,0,1,1,1,1,0),$$
$$ \breve{e}+\breve{h}=\breve{I}={\rm diag}(1,1,1,1,1,1,1,1),$$
$$ \breve{\Omega}^{ab}=gB(-if^{ab8})={\Lambda^2\over2}
\breve{n}^{ab},~~~~~~gB=\Lambda^2\sqrt{3},~~~~~~
\breve{n}^2=\breve{h}.$$
In these notion the operator (\ref{Grgl}) can be written in
the form
\begin{eqnarray}
\label{Grop}
&&(G^{-1})_{\mu\nu}=-\partial_\rho^2P_{\mu\nu}^{(0)}+
(-\breve\nabla_\rho^2+2\Lambda^2)P_{\mu\nu}^{(+)}
+(-\breve\nabla_\rho^2-2\Lambda^2)P_{\mu\nu}^{(-)}
\end{eqnarray}
where the projection operators are introduced
$$ P_{\mu\nu}^{(0)}=\breve{e}\delta_{\mu\nu},~~~~~~
P_{\mu\nu}^{(\pm)}={1\over2}\left[\delta_{\mu\nu}\breve{h}
\pm i\breve{n}b_{\mu\nu}\right],~~~~~~
(P_{\mu\nu}^{(\pm)})^\top=P_{\nu\mu}^{(\pm)}.$$
Here the symbol $"*^\top"$ is connected with color indices.

The equation on the Green function looks like
\begin{eqnarray}
\label{freq}
&&(G^{-1})_{\mu\nu}^{ab}A_\nu^b(x)=
\left[-\breve\nabla_\rho^2\delta_{\mu\nu}-
4i\breve\Omega b_{\mu\nu}\right]^{ab}A_\nu^b(x)\\
&& =\left[-\partial_\rho^2P_{\mu\nu}^{(0)}+
(-\breve\nabla_\rho^2+2\Lambda^2)P_{\mu\nu}^{(+)}
+(-\breve\nabla_\rho^2-2\Lambda^2)P_{\mu\nu}^{(-)}
\right] A_\mu(x)=J_\mu(x).\nonumber
\end{eqnarray}
Then, the Green function can be written in the form
\begin{eqnarray}
\label{GrFC}
&& G_{\mu\nu}(x,y)=F_{\mu\nu}(x-y)+D_{\mu\nu}(x,y),\\
&& F_{\mu\nu}(x-y)=F(Z)P_{\mu\nu}^{(0)},~~~~~~~
F(Z)={1\over-\partial_\rho^2}\delta(z),\nonumber
\end{eqnarray}
\begin{eqnarray*}
D_{\mu\nu}(x,y)&=&e^{i\breve{\Omega}(xby)}D_{\mu\nu}(z),\\
D_{\mu\nu}(z)&=&D^{(+)}(Z)P_{\mu\nu}^{(+)}+
D^{(-)}(Z)P_{\mu\nu}^{(-)},\nonumber\\
&& D^{(\pm)}(Z)={1\over-\breve\nabla_\rho^2(z)\pm2\Lambda^2}
\delta(z),
\end{eqnarray*}
where $z=x-y$ and $Z={\Lambda^2z^2\over4}$.

The free part of the Green function is
\begin{eqnarray}
\label{Free}
&& F(Z)={1\over-\partial_\rho^2}\delta(z)=\left({\Lambda
\over4\pi}\right)^2\cdot{1\over Z}.
\end{eqnarray}

The regular Green function $D^{(+)}$ reads
\begin{eqnarray}
\label{C+}
D^{(+)}(Z)&=&{1\over-\breve\nabla_\rho^2(z)+2\Lambda^2}
\delta(z)=\left({\Lambda\over4\pi}\right)^2
\int\limits_1^\infty du\cdot{u-1\over u+1}\cdot e^{-Zu}\\
&=&\left({\Lambda\over4\pi}\right)^2\left[{e^{-Z}\over Z}
+2e^{Z}{\rm Ei}(-2Z)\right]
\to\left({\Lambda\over4\pi}\right)^2{e^{-Z}\over2Z^2}~~~~~
{\rm for}~~~~Z\to\infty.\nonumber
\end{eqnarray}
It is convenient to represent
\begin{eqnarray}
\label{CC+}
D^{(+)}(Z)&=&\left({\Lambda\over4\pi}\right)^2
\int\limits_1^\infty d{\cal D}_c^{(+)}~e^{-Zc}.
\end{eqnarray}

The operator $-\breve\nabla_\rho^2(z)-2\Lambda^2$
in the function $D^{(-)}(Z)$ contains the so
called zeroth mode, i.e., there exists the eigen
function $\Phi_0$ for which the eigen value equals 0
$$[-\breve\nabla_\rho^2(z)-2\Lambda^2]\Phi_0(z)=0,~~~~~~~
\Phi_0(z)=C_0\cdot e^{-{\Lambda^2\over4}z^2}.$$
In order to find the contribution of this zeroth mode to
the Green function, we should solve the equation
$$[-\breve\nabla_\rho^2(z)-2\Lambda^2]\Psi_0(z)=
J_0\cdot\Phi_0(z)=J_0\cdot e^{-{\Lambda^2\over4}z^2}.$$
This equation can be solved and the behaviour of this
solution is
$$ \Psi_0(z)\sim e^{{\Lambda^2\over4}z^2}~~~~~
{\rm for}~~~~~z^2\to\infty.$$
It means that this function does not belong to the space
${\bf L}^2$ and can not describe any physical state.
Therefore, this function should be excluded from the set
of physical states and the gluon Green function can not
contain this contribution.

Thus, the Green function $D^{(-)}$ without the zeroth mode
looks like
$$ D^{(-)}(Z)={\rm reg}~{1\over-\breve\nabla_\rho^2(z)-
2\Lambda^2}\delta(z)$$
where ${\rm reg}$ means that the zeroth mode should
be excluded. For this aim let us use the resolvent
\begin{eqnarray*}
R(\zeta)&=&{1\over-\breve\nabla_\rho^2(z)-
2\Lambda^2+\zeta}\delta(z)=\int\limits_0^\infty
d\alpha~e^{\alpha\breve\nabla_\rho^2(z)+(2\Lambda^2-\zeta)
\alpha}\delta(z)\\
&=&\int\limits_0^\infty d\alpha~e^{-\zeta\alpha}\left(
{\Lambda^2~e^{\Lambda^2\alpha}\over4\pi\sinh(\Lambda^2\alpha)}
\right)^2\cdot e^{-{z^2\Lambda^2\over4}\coth(\Lambda^2\alpha)}\\
&\to&\left({\Lambda\over4\pi}\right)^2\left\{\int\limits_0^\infty
du~\left[(\coth(u)+1)^2e^{-Z\coth(u)}-4e^{-Z}\right]+
{4\Lambda^2\over\zeta}~e^{-Z}\right\}
\end{eqnarray*}
for $\zeta\to0$.

Then, the Green function $D^{(-)}$ is
\begin{eqnarray}
\label{C-}
D^{(-)}(Z)&=&
\left({\Lambda\over4\pi}\right)^2\int\limits_1^\infty
{du\over u^2-1}\cdot\left[(u+1)^2e^{-Zu}-4e^{-Z}\right]\\
&=&\left({\Lambda\over4\pi}\right)^2\left[{1\over Z}-2{\bf C}
-2\ln(2Z)\right]\cdot e^{-Z}\nonumber\\
&\to&-\left({\Lambda\over4\pi}\right)^2\ln(2Z)e^{-Z}~~~~~
{\rm for}~~~~Z\to\infty.\nonumber
\end{eqnarray}
Here, ${\bf C}=0.577215...$ is the Euler constant.
We shall use the representation
\begin{eqnarray}
\label{CC-}
D^{(-)}(Z)&=&\left({\Lambda\over4\pi}\right)^2
\int\limits_1^\infty d{\cal D}_c^{(-)}~e^{-Zc}.
\end{eqnarray}
\vspace{1.5cm}
Finally, the gluon Green function reads
\begin{eqnarray}
\label{GGG}
&& G_{\mu\nu}(x,y)=F(Z)\cdot P_{\mu\nu}^{(0)}
+e^{i\breve{\Omega}(xby)}D_{\mu\nu}(Z),\\
&& D_{\mu\nu}(Z)=D^{(+)}(Z)\cdot P_{\mu\nu}^{(+)}
+D^{(-)}(Z)\cdot P_{\mu\nu}^{(-)}.\nonumber
\end{eqnarray}

\subsection{Confinement of gluons.}

Confinement problem: four color degrees of freedom are confined,
but the other four color degrees of freedom are not confined.
Our assumption is that the gluon vacuum has a domain structure
${\bf R}=\bigcup\limits_j{\Gamma}_j$ with the scale of distances
$$ R_{hadron}\ll R_{domain}\ll R_{exp}. $$
Here $R_{hadron}\sim{1\over\Lambda}$ is a scale of regions
where the strong quark-gluon-hadron interactions take
place, $R_{domain}$ is a size of domains $\Gamma_j$, $R_{exp}$ is
a macroscopic scale where hadrons are registered. Each domain
is characterized by a direction of the color vector $n^a$ and
this direction is supposed to be random.

Now we would like to give qualitative and semi-quantitative
arguments that this structure of gluon vacuum can provide
for the total gluon confinement if directions of the
color vector $n^a$ are random in each domain. Let us
consider for simplicity the case of group $SU(2)$.
In this case we have the same problem. The $3\times3$
matrix $\breve{n}$ looks like
$$ \breve{n}^{ab}=-i\epsilon^{abc}n^c,~~~~~~(n^cn^c)=1,
~~~~~~(a,b,c=1,2,3) $$
and the eigen values are $\lambda_0=0,~~\lambda_\pm=\pm1$.
The qualitative behaviour of the gluon Green function is
\begin{eqnarray*}
G^{ab}(z)&\sim& \left(e^{-Z\breve{n}}\right)^{ab}\sim
\delta^{ab}e^{-Z}+n^an^b(G_0-e^{-Z})
\end{eqnarray*}
where $G_0$ is the free Green function. For $n=(0,0,1)$ we have
$$ G^{11}(z,n)=G^{22}(z)=e^{-Z},~~~~~~~~G^{33}(z)=G_0,$$
i.e., we have confinement for directions 1 and 2 and
no confinement for direction 3 in the color space.

Now let us consider a gluon which moves in a direction
through $N$ domains. It is clear that the terms with $e^{-Z}$
provide the confinement in each domain but the terms with
$G_0$ permit gluons to travel at large distances. Let a
gluon moves in a direction through $N$ domains. Its amplitude
will be proportional to the factor
$$ A_{fi}^{ab}=n_f^a(n_fn_N)(n_Nn_{N-1})\cdot...\cdot(n_2n_1)
(n_1n_i)n_i.$$
Here $n_j$ is the color vector of the domain $\Gamma_j$. We
can consider $(n_jn_{j-1})=\cos(\theta_j)$ where
$0\leq\theta_j\leq\pi$ is assumed to be a random angle.
Then, one gets
$$ |A_{fi}^{ab}|\leq\prod\limits_{j=1}^N|\cos(\theta_j)|=
\exp\left\{N\int\limits_0^{\pi/2} d\theta\ln\cos(\theta)
\right\}=e^{-Nc}$$
where $c=1.0887...$ . Thus, we see that a large number of domains
can confine gluons.

\section{Two-gluon scalar colorless current}

The matrix element of the second order, which contains
the scalar two-gluon current, looks like
\begin{eqnarray*}
L_2&=&\frac{(ig)^2}{2}\int\!\!\!\int dxdy~
{\rm Tr}([\hat A_\mu(x),\hat A_\nu(x)]\overbrace
{\hat{\cal F}_{\mu\nu}(x))\cdot{\rm Tr}
(\hat{\cal F}_{\alpha\beta}}(y)
[\hat A_\alpha(y),\hat A_\beta(y)])\\
&=&-{(2ig)^2\over2}\int\!\!\! \int dxdy~T_{\mu\nu}^a(x)
\Bigl(-\breve\nabla_\mu^{ab}(x)\breve\nabla_\alpha^{sv}(y)
G_{\nu\beta}^{bv}(x,y)\Bigr)T_{\alpha\beta}^s(y)\\
&=&2g^2\int\!\!\! \int dxdy~T_{\mu\nu}^a(x)
\Bigl[e^{\frac{i}{2}\breve b(x,y)}
{\cal D}_{\mu\alpha,\nu\beta}(x-y)
e^{\frac{i}{2}\breve b(x,y)}\Bigr]^{as}T_{\alpha\beta}^s(y)\\
&=&2g^2\int\!\!\! \int dxdy~{\cal T}_{\mu\nu}^a(x)
\Bigl[e^{\frac{i}{2}\breve b(x,y)}{\cal S}_{\mu\alpha,\nu\beta}(x-y)
e^{\frac{i}{2}\breve b(x,y)}\Bigr]^{as}{\cal T}_{\alpha\beta}^s(y),
\nonumber
\end{eqnarray*}
where
$$ T_{\mu\nu}^a(x)={\rm Tr}~\Bigl([\hat A_\mu(x),\hat A_\nu(x)]
t^a\Bigr),~~~~~~~{\cal T}_{\mu\nu}^a(x)=
{\rm Tr}~\Big(\hat A_\mu(x)\hat A_\nu(x)t^a\Bigr)$$
and
\begin{eqnarray}
&&{\cal D}^{ab}_{\mu\alpha,\nu\beta}(z)=
[\breve\nabla_\mu(z)\breve\nabla_\alpha^{\top}(z)
D_{\nu\beta}(z)]^{ab},\\
&&{\cal S}^{ab}_{\mu\alpha,\nu\beta}(z)=
{\cal D}^{ab}_{\mu\alpha,\nu\beta}(z)+
{\cal D}^{ab}_{\nu\beta,\mu\alpha}(z)-
{\cal D}^{ab}_{\nu\alpha,\mu\beta}(z)-
{\cal D}^{ab}_{\mu\beta,\nu\alpha}(z).\nonumber
\end{eqnarray}

To obtain the function ${\cal D}^{ab}_{\mu\alpha,\nu\beta}(z)$
the following calculations must be fulfiled:
\begin{eqnarray*}
&&-\breve\nabla^{ab}_\mu(x)\breve\nabla^{sv}_\alpha(y)
G^{bv}_{\nu\beta}(x,y)=-\breve\nabla^{ab}_\mu(x)
G^{bv}_{\nu\beta}(x,y)\breve\nabla_\alpha^{\top vs}(y)\\
&&=-\breve\nabla^{ab}_\mu(x)\breve\nabla_\alpha^{\top bv}(y)
G^{vs}_{\nu\beta}(x,y)=
-\breve\nabla^{ab}_\mu(x)\breve\nabla_\alpha^{\top bv}(y)
\left[e^{i\breve b(x,y)} D_{\nu\beta}(x-y)\right]^{vs}\\
&&=-\left[e^{i\breve b(x,y)}\breve\nabla_\mu(x-y)
\breve\nabla_\alpha^{\top}(y-x)
D_{\nu\beta}(x-y)\right]^{as}\\
&&=e^{i\breve
b(x,y)}\breve\nabla_\mu(z)\breve\nabla_\alpha^{\top}(z)
D_{\nu\beta}(z)=e^{{i\over2}\breve b(x,y)}
{\cal D}_{\mu\alpha,\nu\beta}(z)
e^{{i\over2}\breve b(x,y)}.
\end{eqnarray*}
Using the identity
\begin{eqnarray*}
&&t^a_{ij}\Bigl(e^{\frac{i}{2}\breve b(x,y)}\Bigr)^{ab}=
\Bigl(e^{\frac{i}{2}\hat b(x,y)} t^b e^{-\frac{i}{2}\hat
b(x,y)}\Bigr)_{ij},
\end{eqnarray*}
one can obtain
\begin{eqnarray*}
{\rm Tr}~(\hat A_\mu(x)\hat A_\nu(x)t^a)
\left(e^{\frac{i}{2}\breve b(x,y)}\right)^{ab}
&=&{\rm Tr}~(\hat A_\mu(x,y)\hat A_\nu(x,y)t^b) ,\\
\left(e^{\frac{i}{2}\breve b(x,y)}\right)^{rs}
{\rm Tr}~(t^s\hat A_\alpha(y)\hat A_\beta(y))
&=&{\rm Tr}~(t^r\hat A_\alpha(y,x)\hat A_\beta(y,x))
\end{eqnarray*}
where the bilocal gluon fields have the form
\begin{eqnarray*}
&& \hat A_\mu(x,y)=e^{-{i\over2}\hat{b}(x,y)}\hat A_\mu(x)
e^{{i\over2}\hat b(x,y)},~~~~~~~~
\hat A_\alpha(y,x)=e^{-{i\over2}\hat{b}(y,x)}\hat A_\alpha(y)
e^{{i\over2}\hat b(y,x)}
\end{eqnarray*}

So we can rewrite $L_2$ as the follows
\begin{eqnarray}
\label{L2}
&& L_2=2g^2\int\!\!\! \int\!\! dxdy~
{\rm Tr}\Big(\hat A_\mu(x,y)\hat A_\nu(x,y)t^a\Bigr)
{\cal S}_{\mu\alpha,\nu\beta}^{as}(x-y)
{\rm Tr}\Big(t^s \hat A_\alpha(y,x)\hat A_\beta(y,x)\Bigr).
\nonumber\\
\end{eqnarray}

In order to extract the scalar colorless two-gluon state, the
following standard steps should be done:

1. Using the Fierz transformation for the $t^a$-matrices
$$t^a_{ij}
t^b_{kn}=\frac{1}{18}\delta^{ab}\delta^{jk}\delta^{in}+...  ,$$
we extract the four-gluon color singlet
\begin{eqnarray*}
&&{\rm Tr}~\Big(\hat A_\mu(x,y)\hat A_\nu(x,y)t^a\Bigr)
{\cal S}_{\mu\alpha,\nu\beta}^{as}(x-y)
{\rm Tr}~\Big(t^s \hat A_\alpha(y,x)\hat A_\beta(y,x)\Bigr)\\
&&={1\over18}{\rm Tr}~\Big(\hat A_\mu(x,y)\hat A_\nu(x,y)
\hat A_\alpha(y,x)\hat A_\beta(y,x)\Bigr)
{\cal S}_{\mu\alpha,\nu\beta}^{ss}(x-y)+... .
\end{eqnarray*}

2. Next we extract the two-gluon color singlet and octet:
\begin{eqnarray*}
&&{\rm Tr}~\Big(\hat A_\mu(x,y)\hat A_\nu(x,y)
\hat A_\alpha(y,x)\hat A_\beta(y,x)\Bigr)\\
&&=\Bigl(\hat A_\nu(x,y)\hat A_\alpha(y,x)\Bigr)_{ij}
\Bigl(\hat A_\beta(y,x)\hat A_\mu(x,y)\Bigr)_{ji}\\
&&={1\over3}{\rm Tr}~\Big(\hat A_\nu(x,y)\hat
A_\alpha(y,x)\Bigr)\cdot
{\rm Tr}~\Big(\hat A_\beta(y,x)\hat A_\mu(x,y)\Bigr)\\
&&+2{\rm Tr}~\Big(t^a\hat A_\nu(x,y)\hat A_\alpha(y,x)\Bigr)\cdot
{\rm Tr}~\Big(t^a\hat A_\beta(y,x)\hat A_\mu(x,y)\Bigr)
\end{eqnarray*}

3. The two-gluon color singlet can be represented
\begin{eqnarray*}
&& {\cal J}_{\mu\alpha}(x,y)=
{\delta_{\mu\alpha}\over8}{\cal J}(x,y)+
{\cal J}_{\mu\alpha}^{(2)}(x,y),
\end{eqnarray*}
where ${\cal J}(x,y)$ and ${\cal J}_{\mu\alpha}^{(2)}(x,y)$
describes spins zero and two respectively. For spin zero we have
\begin{eqnarray*}
{\cal J}(x,y)&=&2{\rm Tr}~\Big(\hat A_\sigma(x,y)\hat
A_\sigma(y,x)\Bigr)=2{\rm Tr}~\Big(\hat A_\sigma(x)
e^{i\hat b(x,y)}\hat A_\sigma(y)e^{i\hat b(y,x)}\Bigr)\\
&=&A_\sigma^a(x)\Big(e^{i\breve{b}(x,y)}\Bigr)^{ab}
A_\sigma^b(y)=\left(A_\sigma(x)e^{i\breve{b}(x,y)}A_\sigma(y)
\right).
\end{eqnarray*}

4. The effective "potential" connecting two scalar currents is
\begin{eqnarray}
\label{Pot}
{\cal V}(z)&=&{1\over2}{\cal S}^{aa}_{\mu\nu,\nu\mu}(z)=
{\cal D}^{aa}_{\mu\nu,\nu\mu}(z)-{\cal
D}^{aa}_{\mu\mu,\nu\nu}(z)\\
&=&{\rm Tr}~\left[\breve\nabla_\mu(z)\breve\nabla_\nu^{\top}(z)
D_{\nu\mu}(z)-\breve\nabla_\mu(z)\breve\nabla_\mu^{\top}(z)
D_{\nu\nu}(z)\right].\nonumber
\end{eqnarray}

Finally, we get
\begin{eqnarray}
\label{LL2}
&& L_2={g^2\over864}\int\!\!\!\int dx_1dx_2~{\cal J}(x_1,x_2)
{\cal V}(x_1-x_2){\cal J}(x_2,x_1)
\end{eqnarray}

\subsection{The gluon-gluon "potential".}

The gluon-gluon "potential" ${\cal V}(z)$ can be calculated
\begin{eqnarray*}
{\cal V}(z)&=&{\rm Tr}~\left[\breve\nabla_\mu(z)
\breve\nabla_\nu^{\top}(z)D_{\nu\mu}(z)-\breve\nabla_\mu(z)
\breve\nabla_\mu^{\top}(z)D_{\nu\nu}(z)\right].\\
&=&{\rm Tr}~\left[\breve\nabla_\mu(z)P_{\mu\nu}^{(+)}
\breve\nabla_\nu^{\top}(z)
-\breve\nabla_\mu(z)\breve\nabla_\mu^{\top}(z)
P_{\nu\nu}^{(+)}\right]D^{(+)}(Z)\\
&+&{\rm Tr}~\left[\breve\nabla_\mu(z)P_{\mu\nu}^{(-)}
\breve\nabla_\nu^{\top}(z)
-\breve\nabla_\mu(z)\breve\nabla_\mu^{\top}(z)
P_{\nu\nu}^{(-)}\right]D^{(-)}(Z).
\end{eqnarray*}
One can get
\begin{eqnarray*}
&&\left[\breve\nabla_\mu(z)P_{\mu\nu}^{(\pm)}
\breve\nabla_\nu^{\top}(z)D^{(\pm)}(Z)
-\breve\nabla_\mu(z)\breve\nabla_\mu^{\top}(z)
P_{\nu\nu}^{(\pm)}D^{(\pm)}(Z)\right]\\
&=&\breve{h}\left\{{2\over3}\left[-\breve\nabla^2(z)\pm2
\Lambda^2\right]+{\Lambda^2\over2}\left[\mp4-6Z\pm2Z
{d\over dZ}\right]D^{(\pm)}(Z)\right\}.
\end{eqnarray*}
The first term gives $\delta(z)$ and this part of our
potential describes the local interaction which leads to
the standard renormalization and contributions to the
highest perturbation orders. Thus, the potential that is
responsible for the glueball bound state reads
\begin{eqnarray}
\label{PotV}
{\cal W}(z)&=&4\Lambda^2\left\{\left[-2-3Z+Z{d\over dZ}
\right]D^{(+)}(Z)+\left[2-3Z-Z{d\over dZ}\right]D^{(-)}(Z)
\right\}\nonumber\\
&=&\left({\Lambda^2\over2\pi}\right)^2\int\limits_1^\infty
d{\cal W}_a~e^{-Za}.
\end{eqnarray}
It is usefull to introduce the representation
\begin{eqnarray}
\label{WW}
\sqrt{{\cal W}(Z)}&=&{\Lambda^2\over2\pi}
\int\limits_1^\infty d{\cal U}_a~e^{-Za}.
\end{eqnarray}

\section{Nonlocal Currents and Bosonization}

Now let us go to $L_2$ in (\ref{LL2}) and consider the scalar
current $J(x,y)$, which in the center of mass system of two gluons
$\left(x_1=x+{y\over2},~x_2=x-{y\over2}\right)$ becomes
\begin{eqnarray*}
{\cal J}(x,y)&=&\left(A_\sigma\left(x+{y\over2}\right)
e^{i\breve b\left(x+{y\over2},x-{y\over2}\right)}
A_\sigma\left(x-{y\over2}\right)\right)\\
&=&\left(A_\sigma(x)e^{\frac{y}{2}\stackrel{\leftrightarrow}
{\nabla}(x)}A_\sigma(x)\right),
\end{eqnarray*}
where
\begin{eqnarray*}
&&\stackrel{\leftrightarrow}{\nabla}_\mu(x)=
\stackrel{\leftarrow}{\nabla}_\mu(x)-
\stackrel{\rightarrow}{\nabla}_\mu(x),\\
&&\stackrel{\rightarrow}{\nabla}_\mu(x)=
\stackrel{\rightarrow}{\partial}_\mu-i\breve\Omega
b_{\mu\nu}x_\nu,~~~~~~
\stackrel{\leftarrow}{\nabla}_\mu(x)=
\stackrel{\leftarrow}{\partial}_\mu+i\breve\Omega
b_{\mu\nu}x_\nu.
\end{eqnarray*}
Then
\begin{eqnarray*}
L_2&=&{g^2\over864}\int\!\!\!\int dxdy~{\cal J}(x,y)
{\cal W}(y){\cal J}(x,-y)\\
&=&{g^2\over864}\int dx \!\!\int\!\!\!\int dy_1dy_2
\sqrt{{\cal W}(y_1)}{\cal J}(x,y_1)\delta^4(y_1-y_2)
\sqrt{{\cal W}(y_2)}{\cal J}(x,-y_2)
\end{eqnarray*}
Let us introduce an orthonormal system of functions
$\{U_Q(y)\}$
\begin{eqnarray}
\label{orto}
\sum_Q U_Q(y_1) U_Q(y_2)=\delta(y_1-y_2),
~~~~~~~\int dy~U_Q(y) U_{Q\prime}(y)=\delta_{QQ\prime}.
\end{eqnarray}
where $Q$ are quantum numbers. We obtain
\begin{eqnarray}
\label{L2Q}
&&L_2={g^2\over864}\sum_Q \int dx~{\cal J}_Q(x)\cdot
{\cal J}_Q^{+}(x)
\end{eqnarray}
where
\begin{eqnarray*}
&& {\cal J}_Q(x)=\left(A_\sigma(x)V_Q(\stackrel
{\leftrightarrow}{\nabla}(x))A_\sigma(x)\right),\\
&& V_Q(\stackrel{\leftrightarrow}{\nabla}(x))=
\int dy~\sqrt{{\cal W}(y)} U_Q(y)
e^{\frac{y}{2}\stackrel{\leftrightarrow}{\nabla}(x)}.
\end{eqnarray*}
We are interested in the scalar glueball which is described by
the current ${\cal J}_0$, i.e., with the quantum numbers $Q=0$.
The part of $L_2$ with these quantum numbers looks like
\begin{eqnarray}
\label{L20}
&&L_2={g^2_1\over 4}\int dx~{\cal J}_0^2(x),~~~~~~
g^2_1=\frac{g^2}{216}\\
&& {\cal J}_0(x)=\left(A_\sigma(x)V_0(\stackrel{\leftrightarrow}
{\nabla}(x))A_\sigma(x)\right).\nonumber
\end{eqnarray}

The scalar gluon current consists of the confined and non-confined
gluon modes
$$ {\cal J}={\cal J}^{(f)}+{\cal J}^{(c)}=
(A_\mu^{(f)}V_0A_\mu^{(f)})+
(A_\mu^{(c)}V_0A_\mu^{(c)})$$
It should be stressed that this division is $SU(3)$ invariant
because the eigen numbers of the matrix $\breve{n}$ are invariants
of $SU(3)$ group transformations. Therefore,
$$ (A_\mu G^{-1}_{\mu\nu}A_\nu)=
(A_\mu^{(c)}G^{-1}_{\mu\nu}A_\nu^{(c)}(x))+
(A_\mu^{(f)}G^{-1}_{\mu\nu}A_\nu^{(f)}(x)).$$

Our main assumption is that only confined modes lead to a bound
state, but local modes lead to standard renormalizations and
contributions to next perturbation orders. It means that the part
of the partition function, which is responsible for the
glueball formation, reads
\begin{eqnarray*}
Z&=&\int DA^{(c)}~e^{-{1\over2}(A^{(c)}D^{-1}A^{(c)})
+{g^2_1\over 4}({\cal J}_0^{(c)}{\cal J}_0^{(c)})}\\
&=&\int D\varphi e^{-\frac{1}{2}(\varphi\varphi)}
\int DA^{(c)}~e^{-\frac{1}{2}(A^{(c)}D^{-1}A^{(c)})
+{g_1\over\sqrt 2}({\cal J}_0^{(c)}\varphi)}\\
&=&\int D\varphi e^{-\frac{1}{2}(\varphi\varphi)}
\int DA^{(c)}~e^{-\frac{1}{2}(A^{(c)}D^{-1}A_\nu^{(c)})
+{g_1\over\sqrt 2}(A^{(c)}\varphi V_0A^{(c)})}\\
&=&\int D\varphi e^{-\frac{1}{2}(\varphi\varphi)
-{1\over2}{\rm Tr}\ln(1-{\sqrt 2}g_1\varphi V_0D)}.
\end{eqnarray*}
Expanding the logarithm, we extract the quadratic part over
$\varphi(x)$ in the form
\begin{eqnarray*}
Z&=&\int D\varphi e^{-\frac{1}{2}(\varphi{\cal H}^{-1}
\varphi)+O(\phi^3)}.
\end{eqnarray*}
where
\begin{eqnarray*}
&& {\cal H}^{-1}(x_1-x_2)=\delta(x_1-x_2)-g_1^2\Pi(x_1-x_2),
\end{eqnarray*}
\begin{eqnarray}
&&\Pi(x_1-x_2)={\rm Tr}~
\Bigl\{V_0(\stackrel{\leftrightarrow}{\nabla}_{x_1})D(x_1,x_2)
V_0(\stackrel{\leftrightarrow}{\nabla}_{x_2})D(x_2,x_1)\Bigr\}.
\nonumber
\end{eqnarray}
The polarization operator $\Pi(z)$ can be written as follows
$(z=x_1-x_2)$
\begin{eqnarray}
\label{Pol}
&&\Pi(z)=\int\!\!\!\int dy_1 dy_2~\sqrt{{\cal W}(y_1)}
U_0(y_1)\Pi(z|y_1,y_2)\sqrt{{\cal W}(y_2)}U_0(y_2)
\end{eqnarray}
where
\begin{eqnarray*}
&& \Pi(z|y_1,y_2)\\
&&={\rm Tr}~\left\{e^{\frac{y_1}{2}\stackrel{\leftrightarrow}
{\nabla}(x_1)}e^{i\breve\Omega(x_1bx_2)}D_{\alpha\beta}(x_1-x_2)
e^{\frac{y_2}{2}\stackrel{\leftrightarrow}{\nabla}(x_2)}
e^{i\breve\Omega(x_2bx_1)}D_{\beta\alpha}^\top(x_2-x_1)\right\}\\
&&={\rm Tr}~\left\{e^{\frac{y_1}{2}\stackrel{\leftrightarrow}
{\nabla}(z)}D_{\alpha\beta}(z)e^{\frac{y_2}{2}\stackrel
{\leftrightarrow}{\nabla}(z)}D_{\beta\alpha}^\top(z)\right\}\\
&&={\rm Tr}~\left\{e^{i\breve\Omega((y_1-y_2)bz)}
D_{\alpha\beta}\left(z-{y_1+y_2\over2}\right)
D_{\beta\alpha}^\top\left(-z-{y_1+y_2\over2}\right)\right\}\\
&&=2{\rm Tr}\left[\breve{h}e^{i\breve\Omega((y_1-y_2)bz)}
\sum\limits_{j=\pm}D^{(j)}\left(Z_{(+)}\right)
D^{(j)}\left(Z_{(-)}\right)\right]
\end{eqnarray*}
with
$$  Z_{(\pm)}={\Lambda^2\over4}
\left(z\pm{y_1+y_2\over2}\right)^2.$$

\subsection{Polarization operator}

We do not know the eigen function $U_0(y)$; so
we shall use the variation method with the test function
\begin{eqnarray*}
&&U_0(y,\gamma)={\Lambda^2\over2\pi}\cdot\ e^{-\gamma Y},
~~~~~~Y={\Lambda^2\over4}y^2,
\end{eqnarray*}
where $\gamma$ is a variation parameter. Our point of view
is that this function should be a quite good approximation for
the ground state of the Bethe-Salpeter equation.

The Fourier transfrmation of the polarization operator looks like
\begin{eqnarray}
\label{PolF}
&&\tilde{\Pi}(p^2,\gamma)=\int dz~e^{ipz}\Pi(z)
={1\over\pi^2}\left({\Lambda^2\over4\pi}\right)^6\cdot R(p^2,\gamma)\\
&& R(p^2,\gamma)=\gamma^2\int\!\!\!\!\int\limits_1^\infty
d{\cal U}_{a_1}d{\cal U}_{a_2}\sum\limits_{j=\pm}
\int\!\!\!\!\int\limits_1^\infty d{\cal D}_{c_1}^{(j)}
d{\cal D}_{c_2}^{(j)}~I(a_1,a_2,c_1+\gamma,c_2+\gamma;p^2).
\nonumber
\end{eqnarray}
\begin{eqnarray*}
&&I(a_1,a_2,c_1,c_2;p^2)={\rm Tr}~2\breve{h}
\int dz\int\!\!\!\int dy_1 dy_2\\
&&\cdot e^{ipz-a_1{y_1^2\Lambda^2\over4}-a_2{y_2^2\Lambda^2\over4}
-c_1{\Lambda^2\over4}\left(z-{y_1+y_2\over2}\right)^2
-c_2{\Lambda^2\over4}\left(z+{y_1+y_2\over2}\right)^2
+i\breve\Omega((y_1-y_2)bz)}\\
&&=\left({4\pi\over\Lambda^2}\right)^6\cdot
{8\over[K(a_1,a_2,c_1,c_2)]^2}\cdot e^{-{p^2\over\Lambda^2}
\cdot{k(a_1,a_2,c_1,c_2)\over K(a_1,a_2,c_1,c_2)}},\\
&& k(a_1,a_2,c_1,c_2)=a_1a_2+{1\over4}(a_1+a_2)(c_1+c_2),\\
&& K(a_1,a_2,c_1,c_2)=(a_1+a_2)(1+c_1c_2)+(c_1+c_2)(1+a_1a_2).
\end{eqnarray*}
Finally, we have
\begin{eqnarray}
\label{pol}
&& g_1^2\tilde{\Pi}(p^2,\gamma)\\
&&={\alpha_s\over\pi}\cdot{4\gamma^2
\over27}\int\!\!\!\!\int\limits_1^\infty d{\cal U}_{a_1}
d{\cal U}_{a_2}\sum\limits_{j=\pm}\int\!\!\!\!\int\limits_1^\infty
d{\cal D}_{c_1}^{(j)}d{\cal D}_{c_2}^{(j)}
\cdot{\exp\left\{-{p^2\over\Lambda^2}
\cdot{k\left(a_1,a_2,c_1+\gamma,c_2+\gamma\right)
\over K\left(a_1,a_2,c_1+\gamma,c_2+\gamma\right)}\right\}
\over[K\left(a_1,a_2,c_1+\gamma,c_2+\gamma\right)]^2}\nonumber
\end{eqnarray}
where $\alpha_s={g^2\over4\pi}$. Next step is to calculate
\begin{eqnarray*}
&& \max\limits_\gamma~g_1^2\tilde{\Pi}\left(-{M^2\over\Lambda^2},
\gamma\right)={\alpha_s\over\pi}\cdot\tilde{\Pi}\left(
{M^2\over\Lambda^2}\right)
\end{eqnarray*}
and the mass of scalar glueball is defined by the equation
\begin{eqnarray}
\label{alM}
&&{\alpha_s\over\pi}\cdot\tilde{\Pi}\left({M^2\over\Lambda^2}\right) =1
\end{eqnarray}
The results of calculations are given in the Table.

\begin{tabular}{|c|c|c|}
\hline
   &  & \\
${M\over\Lambda}$ & $\alpha_s$ & $M$ \\
 & & $(\Lambda=446~Mev)$ \\
 & &\\
\hline
& &\\
3.0 & 1.5120 & 1337\\
3.1 & 1.0609 & 1381 \\
3.2 & 0.7369 & 1426\\
3.3 & 0.5067 & 1470 \\
& &\\
3.4 & 0.3450 & 1515 \\
3.5 & 0.2325 & 1560 \\
3.6 & 0.1551 & 1604 \\
&  & \\
3.7 & 0.1024 & 1649 \\
3.8 & 0.0669 & 1694 \\
3.9 & 0.0432 & 1738 \\
4.0 & 0.0277 & 1782 \\
    & & \\
\hline
\end{tabular}

\vspace{1cm}
The gluon condensate
\begin{eqnarray*}
&& \left\langle G_{\mu\nu}^aG_{\mu\nu}^a\right\rangle=
B_{\mu\nu}^aB_{\mu\nu}^a=n^an^aB^2b_{\mu\nu}b_{\mu\nu}=
4B^2,\\
&& \Lambda^4={1\over3}g^2B^2=
{\pi^2\over3}\cdot{g^2\over4\pi^2}\cdot4B^2=
{\pi^2\over3}\cdot{\alpha_s\over\pi}\cdot
\left\langle G_{\mu\nu}^aG_{\mu\nu}^a\right\rangle=
{\pi^2\over3}\langle{\cal O}_G\rangle,\\
&&\langle{\cal O}_G\rangle=.012x_G~(Gev)^4,\\
&& \Lambda=446x_G^{.25}~Mev,
\end{eqnarray*}
where $x_G\approx1$ (see, for example, \cite{shif}).
The values of $\alpha_s(\Lambda^2)$ are given in \cite{hinch}.

\subsection{Conclusion}

In the conclusion we would like to list the main points of
our approach.
\begin{itemize}
\item Analytic confinement plus weak coupling lead to
the Regge spectrum of bound states.
\item Gluon self-dual vacuum field $\hat{B}_\mu(x)$ leads
to full quark confinement and partial gluon confinement.
\item The field $\hat{B}_\mu(x)$ has directions in color
and configuration spaces; so the domain structure of the
whole space ${\bf R}^4$ with appropriate averaging over
these directions is needed to provide full gluon
confinement and relativistic covariance.
\item Gluon propagator in the field $\hat{B}_\mu(x)$ has
the zeroth modes, which are not the Goldstone degrees of
freedom. We exclude them, because they lead to a
nonphysical growth of the propagator in the $x$-space for
$x\to\infty$.
\item Numerical results are quite reasonable and
with the results of \cite{mod1,mod2,mod3} give hope
that our main assumption about $\hat{B}_\mu(x)$
contains "a grain of truth".
\end{itemize}

\subsection{Appendix}

We use the following method of approximations.
Let a positive function
\begin{eqnarray}
&& f(t)=\int\limits_1^\infty d\sigma_u~e^{-tu}
\end{eqnarray}
be given, where $d\sigma_u$ is a positive measure.

Our aim is to get the approximation
\begin{eqnarray}
&& f(t)\approx f_a(t)=\sum\limits_{j=1}^n A_j~e^{-a_jt}
\end{eqnarray}
on the interval $0\leq t_i\leq t\leq t_f\leq\infty$.

Our procedure is the following. Let us choose $n$ points
$$ \{t_1,t_2,...,t_n\},~~~~~t_j\in[t_i,t_f]$$
and calculate
$$ a_j=-{f'(t_j)\over f(t_j)},~~~~~~~~~(j=1,...,n).$$
Then, we solve the system of equations
\begin{eqnarray*}
&& f(t_i)=\sum\limits_{j=1}^n A_j~e^{-a_jt_i},~~~~~~~(i=1,...,n)
\end{eqnarray*}
so that we get the numbers
$$ \{A_1,...,A_n\}$$
and we have the approximation
\begin{eqnarray}
&& f(t)=\int\limits_1^\infty d\sigma_u~e^{-tu}
\approx f_a(t)=\sum\limits_{j=1}^n A_j~e^{-a_jt}.
\end{eqnarray}
The accuracy of this approximation can be evaluated by
\begin{eqnarray*}
&& \delta={||f-f_a||\over||f||},~~~~~~
||f||=\int\limits_{t_i}^{t_f} dt|f(t)|
\end{eqnarray*}
The accuracy depends on a particular choice of the points
$\{t_1,...,t_n\}$.

In our case, we have

\vspace{1cm}
\begin{tabular}{|c|c|c|c|c|c|c|}\hline
 &\multicolumn{2}{|c|}{} &
\multicolumn{2}{|c|}{} &\multicolumn{2}{|c|}{}\\
 &\multicolumn{2}{|c|}{$\sqrt{{\cal W}(t)}$} &
\multicolumn{2}{|c|}{$D^{(+)}(t)$} &
\multicolumn{2}{|c|}{$D^{(-)}(t)$}\\
 &\multicolumn{2}{|c|}{} &
\multicolumn{2}{|c|}{} &\multicolumn{2}{|c|}{}\\
\hline
$t_j$ & \multicolumn{2}{|c|}{$.04,~.4,~1,~2.2,~4,~7,~10$} &
\multicolumn{2}{|c|}{$.01,~.05,~.2,~.4,~1.2$} &
\multicolumn{2}{|c|}{$.01,~.05,~.1,~.3,~.5,~.9,~3$}\\
\hline
j & $A_j$ & $a_j$ & $A_j$ & $a_j$ & $A_j$ & $a_j$ \\
\hline
 &    &  &  &   &   & \\
1 &  7.009  & 3.341 & 179.4 &  106.3 & 175.2  & 96.62\\
2 &  0.485  & 2.227 & 29.42 &  24.04 & 36.16  & 19.76\\
3 &  -8.234 & 0.959 & 6.334 &  7.607 & 152.7  & 10.94\\
4 &  -1.532 & 0.002 & 0.281 &  4.592 & 27.77  & 6.554\\
5 &  55.497 & 0.144 & 1.031 &  2.365 & -171.8 & 10.46\\
6 &  -257.1 & 0.197 &       &        & 3.971  & 1.836\\
7 &  208.3  & 0.215 &       &        & -2.783 & 0.823\\
 &    &  &  &   &   & \\
\hline
$\delta$ &\multicolumn{2}{|c|}{$0.041$} &
\multicolumn{2}{|c|}{$0.027$} &
\multicolumn{2}{|c|}{$0.035$}\\
\hline
\end{tabular}


\begin{thebibliography}{99}
\bibitem{close} F.E.Close. "Glueballs: A central mystery",
hep-ph/0006288 26.06.00;
\bibitem{ellis} J.Ellis, H.Fujii, D.Kharzeev. "Scalar Glueball-Quarkonium Mixing and Structure
of QCD Vacuum". hep-ph/9909322 10.09.99;
\bibitem{sim} A.B.Kaidalov, Yu.A.Simonov, "Glueball masses and Pomeron
trajectory in nonperturbative QCD approach", hep-ph/9912434 (2000),
\bibitem{forkel} H.Forkel, "Scalar Gluonium and Instantons",
hep-hp/0005004 (2000);
\bibitem{mod1} G V Efimov, S.N. Nedelko, Phys. Rev. D51 (1995) 174;
 Eur. Phys. J. C1 (1998) 343;
\bibitem{mod2} Ja. V Burdanov, G V Efimov, S.N. Nedelko, S.A. Solunin,
   Phys. Rev. D54 (1996) 4483;
\bibitem{mod3} G V Efimov, A.C. Kalloniatis, S.N. Nedelko,
   Phys. Rev. D59 (1999) 014026;
\bibitem{efiv} G.V.Efimov, M.A.Ivanov, "The Quark Confinement Model
of Hadrons", IOP Publishing Ltd, London, 1993;
\bibitem{efim} G.V. Efimov, "Bound states in QFT, scalar fields."
 hep-ph/9907483 (1999);
\bibitem{shif} M.A.Shifman, ITEP Lectures on Particle Physics and
Field Theory, Vol I, World Scientific Lecture Notes, World Scientific-
Vol 62, Singapure, 1999;
\bibitem{hinch} I.Hinchliffe. "The QCD Coupling Constant".
hep-ph/0004186 19.04.00.
\end{thebibliography}
\end{document}